\documentclass[a4paper,11pt]{article}
\usepackage{pos}
\usepackage[english]{babel}           
\usepackage{color}                    
\usepackage{titlesec}                 
\usepackage{newlfont}          	      
\usepackage[T1]{fontenc}              

\usepackage{tikz}                     
\usepackage{tikz-feynman}             
\tikzfeynmanset{compat=1.1.0}         
\usetikzlibrary{bending}              
\usetikzlibrary{decorations.markings} 
\usetikzlibrary {arrows.meta}         
\usetikzlibrary{snakes}               
\usepackage{graphicx}                 
\usepackage{pdfpages}                 
\usetikzlibrary{decorations.pathreplacing}               

\usepackage{braket}
\usepackage{amsmath}                  
\usepackage{dsfont}                   
\usepackage{simpler-wick}             

\pgfdeclaredecoration{Snake}{initial}
{\state{initial}[switch if less than=+.625\pgfdecorationsegmentlength to final,
                  width=+.3125\pgfdecorationsegmentlength,
                  next state=up]{
    \pgfpathmoveto{\pgfqpoint{0pt}{\pgfdecorationsegmentamplitude}}
  }
  \state{down}[switch if less than=+.8125\pgfdecorationsegmentlength to end down,
               width=+.5\pgfdecorationsegmentlength,
               next state=up]{
    \pgfpathcosine{\pgfqpoint{.25\pgfdecorationsegmentlength}{-1\pgfdecorationsegmentamplitude}}
    \pgfpathsine{\pgfqpoint{.25\pgfdecorationsegmentlength}{-1\pgfdecorationsegmentamplitude}}
  }
  \state{up}[switch if less than=+.8125\pgfdecorationsegmentlength to end up,
             width=+.5\pgfdecorationsegmentlength,
             next state=down]{
    \pgfpathcosine{\pgfqpoint{.25\pgfdecorationsegmentlength}{\pgfdecorationsegmentamplitude}}
    \pgfpathsine{\pgfqpoint{.25\pgfdecorationsegmentlength}{\pgfdecorationsegmentamplitude}}
  }
  \state{end down}[width=+.3125\pgfdecorationsegmentlength,
                   next state=final]{
     \pgfpathcosine{\pgfqpoint{.25\pgfdecorationsegmentlength}{-1\pgfdecorationsegmentamplitude}}
     \pgfpathsine{\pgfqpoint{.25\pgfdecorationsegmentlength}{-1\pgfdecorationsegmentamplitude}}
  }
  \state{end up}[width=+.3125\pgfdecorationsegmentlength,
                 next state=final]{
     \pgfpathcosine{\pgfqpoint{.25\pgfdecorationsegmentlength}{1\pgfdecorationsegmentamplitude}}
     \pgfpathsine{\pgfqpoint{.25\pgfdecorationsegmentlength}{1\pgfdecorationsegmentamplitude}}
  }
  \state{final}{\pgfpointdecoratedpathlast}
}

\newlength{\wdth}

\newdimen\wickgap \wickgap=2pt 
\newbox\wickbox
\def\doublewick\c#1#2\c#3{\setbox\wickbox=\hbox{$#2$}
 \c#1{\phantom{\kern-.5\wickgap\copy\wickbox}}
 \kern-\wd\wickbox\kern.5\wickgap
 \rlap{\copy\wickbox}
 \c#3{\phantom{\kern.5\wickgap\box\wickbox}}
 \kern-.5\wickgap}

\title{An Update on the Isospin-Breaking Effects in the Pion Decay Constant with Staggered Quarks}
\ShortTitle{An Update on the IBE in the Pion Decay Constant with Staggered Quarks}

\author*[a]{Alessandro Cotellucci}
\author[a,b]{Davide Giusti}

\onbehalf{on behalf of the Budapest-Marseille-Wuppertal collaboration}

\affiliation[a]{Jülich Supercomputing Centre, Forschungszentrum Jülich, D-52428 Jülich, Germany}
\affiliation[b]{Fakult\"at f\"ur Physik, Universit\"at Regensburg, 93040, Regensburg, Germany}
\emailAdd{a.cotellucci@fz-juelich.de}

\abstract{We present an update on the ongoing computation of the isospin-breaking effects in the Pion Decay Constant from the BMW Collaboration. The calculation is carried out with N$_f$=2+1+1 staggered quarks with a near-physical pion mass and QED$_{\text{L}}$. We give an update on the isosymmetric value and the current determination used to compute the gradient-flow scale $w_{0}$, then we present some preliminary results of the valence-valence contribution to the axial-pseudoscalar correlator for different volumes and lattice spacings. We also discuss the next steps and plans.}

\FullConference{The 42nd International Symposium on Lattice Field Theory (LATTICE2025)\\
2-8 November 2025\\
Tata Institute of Fundamental Research, Mumbai, India\\}

\tikzfeynmanset{warn luatex=false}
\begin{document}
\maketitle
\section{Motivations}
\begin{figure}[h!]
  \centering
  \includegraphics[width=10cm]{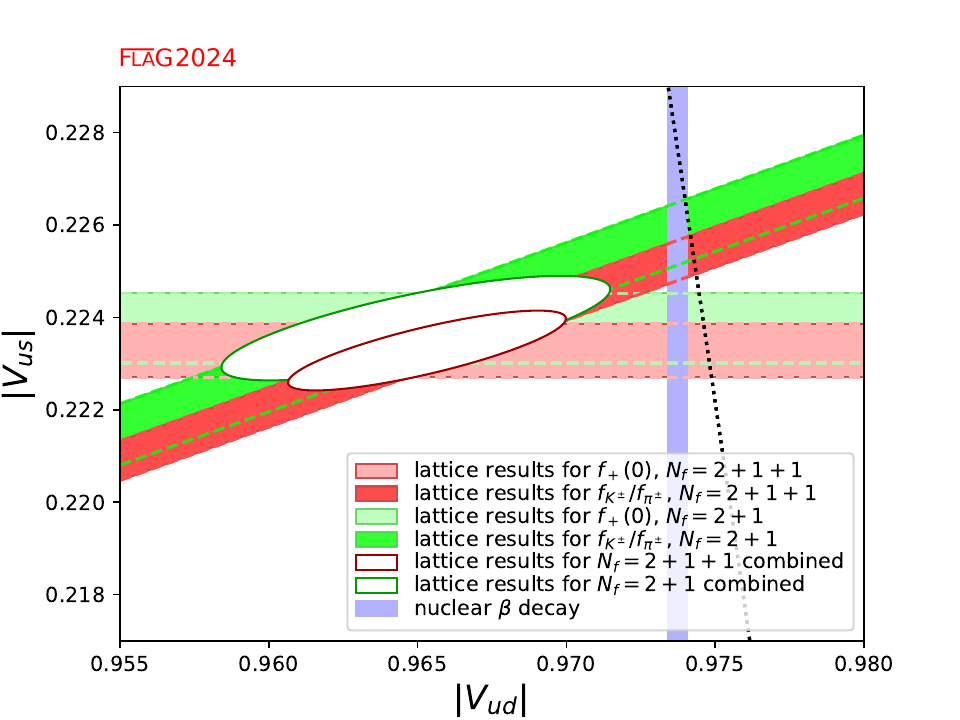} 
    \caption{\label{fig:FLAG2024}Status of the determination of the $V_{ud}$ and $V_{us}$ elements  of the CKM matrix as reported by the Flavour Lattice Averaging Group~\cite{FlavourLatticeAveragingGroupFLAG:2024oxs}.}
\end{figure}

The unitarity of the first row of the CKM matrix is an open problem in precision flavour physics, as shown in Figure~\ref{fig:FLAG2024} from~\cite{FlavourLatticeAveragingGroupFLAG:2024oxs}. At a current level of precision, isospin-breaking effects become relevant.

In Isosymmetric QCD the pion decay constant $f_{\pi}$ is defined by the hadronic matrix element:
\begin{equation}
\bra{0}\bar{u}\gamma_{4}\gamma_{5}d\ket{\pi(\vec{0})}=M_{\pi}f_{\pi}.
\end{equation}
To compute the pion decay constant in QCD+QED $F_{\pi}$ one needs to consider the leptonic decay rate $\pi^{+}\rightarrow l^{+}\nu_{l}(\gamma)$, in the PDG parametrization it is:
\begin{equation}
F_{\pi}^{2}=\frac{\Gamma_{\pi^{+}\rightarrow l^{+}\nu_{l}(\gamma)}}{\frac{G_{\text{F}}^{2}}{8\pi}|V_{ud}|^2M_{\pi}m_{l}^2\left(1-\frac{m_{l}^2}{M_{\pi}^2}\right)^{2}}=f_{\pi}^2\left[1+\delta R_{\pi}\right]
\end{equation}
with $\delta R_{\pi}$ the isospin-breaking correction to the decay constant. The full computation of $\delta R_{\pi}$ requires the renormalization of the electroweak hamiltonian:
\begin{equation}
\mathcal{H}_{W}= \frac{G_{\text{F}}}{\sqrt{2}}V_{ud}^{*}\left[\bar{d}\gamma^{\mu}\left(1-\gamma_{5}\right)u\right]\left[\bar{\nu_{l}}\gamma_{\mu}\left(1-\gamma_{5}\right)l\right].
\end{equation}
There is one lattice computation of the isospin-breaking corrections in the pion and kaon decay constants by the RM123-Southampton collaboration~\cite{DiCarlo:2019thl}. In the ratio $f_{K}/f_{\pi}$ the renormalization constant cancel (in a massless scheme) so there are more lattice computations of the correction to the latter, from the RM123-Southampton collaboration~\cite{Giusti:2017dwk}, from RBC-UKQCD~\citep{Boyle:2022lsi} and a more recent from RBC-UKQCD with QED$_{\infty}$~\citep{Christ:2025ufc}.

The pion decay constant $F_{\pi}$ is used for the scale setting in the BMW/DMZ determination of the anomalous magnetic moment of the muon~\cite{BMW:2026qvl}.

\section{Mixed Determination}\label{sec:2_mixed_det}
The isospin-breaking effects in the pion decay constant are computed by combining the work from the RM123-Southampton collaboration~\cite{DiCarlo:2019thl} and the BMW results for isosymmetric QCD and the sea quark IBE.

The dependency of the pion decay constant in units of $w_{0}$ on the quark electric charge and bare quark masses is parametrized as follows (up to higher order effects):
\begin{equation}
\begin{aligned}
w_{0} F_{\pi} = & A + B F_{\pi}^{-2}M_{\pi^{+}}^{2}+ C F_{\pi}^{-2}\left(M_{K^{\pm}}^{2}+M_{K^{0}}^{2}-M_{\pi^{\pm}}^{2}\right)/2 \\
& + E e_{v}^2 + F e_{v}e_{s}+Ge_{s}^2
\end{aligned}\label{eq:Fpi_par}
\end{equation}
where $e_{v}$ and $e_{s}$ are, respectively, the valence and sea quark charges. The coefficients $B$ and $C$ are mass derivatives and are computed in QCD. The coefficients $E$, $F$ and $G$ are the valence-valence, sea-valence and sea-sea derivatives. 

Once all the coefficients in Eq.~\eqref{eq:Fpi_par} are determined, the parametrization can be split into the QCD+seaQED and the valQED part:
\begin{equation}
\left[w_{0} F_{\pi}\right]_{\text{QCD+QED}}=\left[w_{0}F_{\pi}\right]_{\text{QCD+seaQED}}+\left[w_{0}F_{\pi}\right]_{\text{valQED}},
\end{equation}
both of the terms on the right-hand side are dependent on the scheme used to define QCD. It is fundamental to remember that $w_{0}$ has not valence quak isopsin-breaking effects so $\left[w_{0}F_{\pi}\right]_{\text{valQED}}$ includes only the valence quark IBE on $F_{\pi}$. The quantity $\left[w_{0}F_{\pi}\right]_{\text{QCD+seaQED}}$ is determined with the BMW data, but the valence contribution is computed from the RM123-Southampton result\footnote{The valence QED contribution to the pion decay rate from Ref.~\cite{DiCarlo:2019thl} is given in a GRS scheme in which the quark masses, renormalized at $\mu = 2$ GeV in $\overline{\text{MS}}$, remain constant as the electromagnetic coupling is turned on or off and the overall lattice scale is fixed using $f_\pi$. In the electroquenched approximation adopted there valence QED corrections do not affect the renormalization of the strong coupling -- \textit{i.e.} the GRS condition $[\widehat{g_s}]_{\text{QCD}}(\mu) = [\widehat{g_s}]_{\text{QCD+valQED}}(\mu)$ $(\ast)$, where $\ \widehat{}\ $ indicates the quantities renormalized in the respective theories, holds trivially -- and thus do not change the lattice spacing from its pure QCD value at fixed $\beta$. We stress that in the full QCD+QED calculation presented here the lattice scale is consistently set using $F_{\pi}$ and, since the purely gluonic quantity $w_{0}$ is free of valence QED contributions -- \textit{viz.} a relation similar to $(\ast)$ occurs for $w_{0}$ as well --, it is justified to use the value of $\delta R_{\pi}$ from Ref.~\cite{DiCarlo:2019thl} as input in our computation to evaluate the term $[w_{0} F_{\pi}]_{\text{QCD+valQED}} / [w_{0} F_{\pi}]_{\text{QCD}}$ and Eq.~\eqref{eq:fpi_comb} without introducing scheme ambiguities.}.

To combine the two results, the same scheme has to be employed. The work~\cite{DiCarlo:2019thl} is defined in the GRS scheme by the quantities~\cite{Giusti:2017dmp}:
\begin{equation}
\begin{aligned}
\left[M_{\pi}\right]^{\text{GRS}} = & 135.0(2)\ \text{MeV}\\
\left[\frac{1}{2}\left(M_{K^{\pm}}+M_{K^{0}}\right)\right]^{\text{GRS}} = & 494.6(1)\ \text{MeV}\\
\left[f_{\pi}\right]^{\text{GRS}} = & 130.65(12)\ \text{MeV} \\
\left[\delta R_{\pi}\right]^{\text{GRS}} = & 0.0153(19)\footnotemark.
\end{aligned}\footnotetext{The value used for our determination of $w_{0}F_{\pi}$ is $\delta R_{\pi}=0.0150(18)$, it does not include the original $\chi$PT estimate of the sea-quark IBE.}
\end{equation}
Since they were working in the electroquenched approximation, which neglects the sea quark IBE, the QCD result includes the sea quark isospin-breaking effects following:
\begin{equation}
\left[*\right]^{\text{GRS}} = \left[*\right]_{\text{QCD+QED}}^{\text{phys}}-\left[*\right]_{\text{valQED}}^{\text{GRS}}= \left[*\right]^{\text{GRS}}_{\text{QCD+seaQED}}.
\label{eq:grstoqcdseaqed}
\end{equation}
Using the parametrization \eqref{eq:Fpi_par}, the identity \eqref{eq:grstoqcdseaqed} can be used to evaluate $\left[w_{0}F_{\pi}\right]^{\text{GRS}}_{\text{QCD+seaQED}}$ which can then be combined with the RM123S results using the relation, up to negligible $O\left(\alpha_{\text{em}}^2\right)$ corrections:
\begin{equation}
\left[w_{0}F_{\pi}\right]_{\text{QCD+QED}} = \left[w_{0}F_{\pi}\right]_{\text{QCD+seaQED}}^{\text{GRS}}\sqrt{1+\left[\delta R_{\pi}\right]^{\text{GRS}}}.
\label{eq:fpi_comb}
\end{equation}

\section{Isosymmetric QCD value in the FLAG scheme}\label{sec:Iso_QCD}
\begin{figure}[h!]
\centering
\includegraphics[width=8cm]{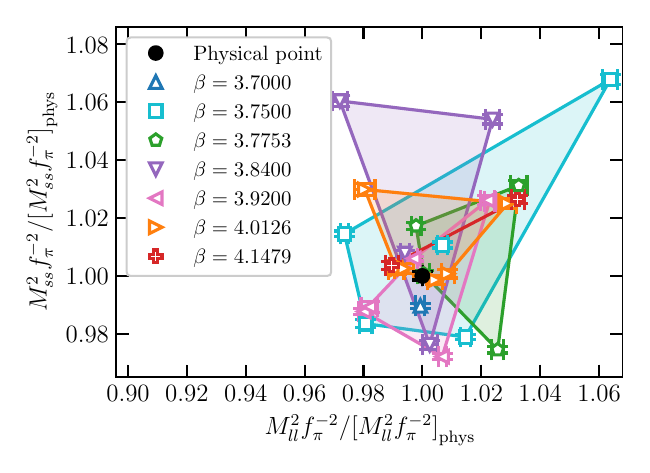}
 \caption{\label{fig:ens_isoqcd} Landscape of the ensembles used for the isoQCD determination of $w_{0}f_{\pi}$ around the physical point from~\cite{BMW:2026qvl}.} 
\end{figure}

The isosymmetric QCD value of $w_{0}f_{\pi}$ is computed using the ensembles of the BMW collaboration shown in Figure~\ref{fig:ens_isoqcd} with QCD tree-level Symanzik action for the gauge field and N$_{f}$=2+1+1 Staggered fermions coupled with 4 levels of stout smearing with radius $\rho=0.125$. The decay constant is computed from the pseudoscalar two-point function through the chiral WTI. 

\begin{figure}[!ht]
\centering
\hspace{-0.2cm}\includegraphics[width=9cm]{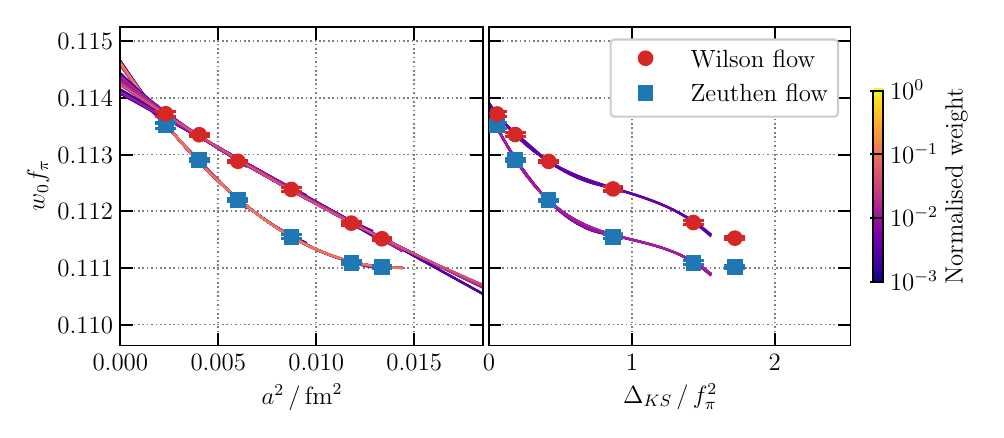}\\
\includegraphics[width=7cm]{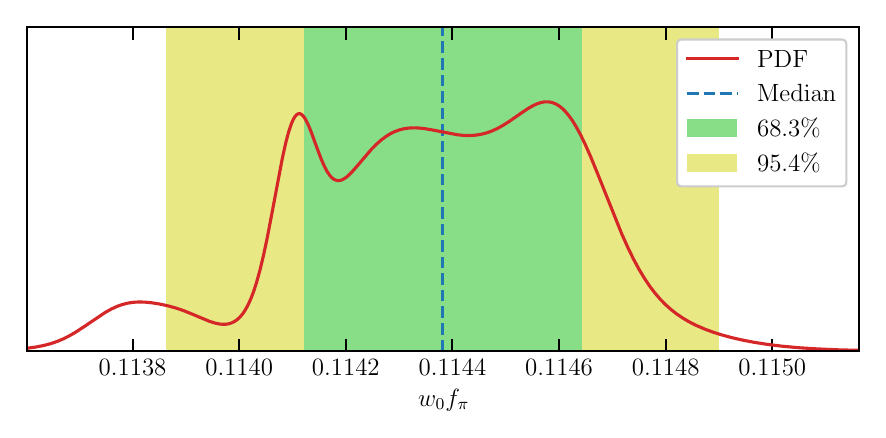}\\
 \caption{\label{fig:cont_lim_isoqcd} Continuum limit extrapolation of the iso QCD value of $w_{0}f_{\pi}$. The upper plot represents the continuum extrapolation both in the lattice spacing and in the taste-symmetry breaking terms $\Delta_{KS}$ (computed by comparing the masses of different meson
tastes on our ensembles); in the lower plot the probability distribution function (PDF) including both statistical and systematic variations is shown
in red.} 
\end{figure}

The analysis follows the strategy presented in~\cite{Boccaletti:2024guq}. To have a better control over the continuum extrapolation we add the determination of $w_{0}$ using the Zeuthen-flow~\cite{Ramos:2015baa} in addition to the Wilson-flow~\cite{Luscher:2010iy}, Figure~\ref{fig:cont_lim_isoqcd} shows the continum extrapolation of $w_{0}f_{\pi}$ in the lattice spacing and in the taste-breaking term and the relative probability distribution function of both the statistical and sistematic effect. The result is presented in the FLAG scheme~\cite{FlavourLatticeAveragingGroupFLAG:2021npn}.

Table~\ref{fig:err_budg_isoqcd} shows the breakdown of the total error on $w_{0}f_{\pi}$. The uncertainty is dominated by the systematic error and, in particular, the biggest contribution is represented by the continuum extrapolation under the entries: \textit{Type of gradient flow}, \textit{Lattice spacing cuts} and \textit{Order of fit polynomials}. To improve the determination, we are generating data at a finer value of the lattice spacing. Other sources of systematic error listed are: \textit{Pseudoscalar fits} (effects of the fit range of the effective pion decay constant) and \textit{Finite volume $\chi$PT order} (the difference in the finite volume effects between NLO and NNLO $\chi$PT).

\begin{table}[!ht]
	\small
    \centering
	\begin{tabular}{l|r|r}
    Median & \multicolumn{2}{c}{0.11438}\\
    \hline
    Total error & 0.00026 & 0.22736 \% \\
    Statistical error &  0.00012 & 0.10786 \% \\
    Systematic error &  0.00023 & 0.20015 \% \\
    \hline
    Type of gradient flow & 0.00013 & 0.10966 \% \\
    Pseudoscalar fits & 0.00004 & 0.03144 \% \\
    Finite volume $\chi$PT order & 0.00001 & 0.01023 \% \\
    Lattice spacing cuts & 0.00006 & 0.04956 \% \\
    Order of fit polynomials & 0.00019 & 0.16215 \% \\
	\end{tabular}
 	\caption{\label{fig:err_budg_isoqcd}Error budget of the isoQCD value of $w_{0}f_{\pi}$. The first row contains the median value, from the second on the errors are presented with the value and the percentage relative value. The last five rows present a breakdown of the systematic error on $w_{0}f_{\pi}$.} 
\end{table} 

The isosymmetric part of $w_{0}f_{\pi}$ is determined with a $0.23\%$ precision, thus making it necessary to include IBE.

\section{Sea Quark Isospin-Breaking Effects}
As mentioned in Section~\ref{sec:2_mixed_det}, the sea quark isospin-breaking effects to $w_{0}F_{\pi}$ are determined using the BMW ensemble. The sea-sea diagrams contributing to the electromagnetic derivative of the pion decay constant are:
\begin{equation}
\vcenter{\hbox{\begin{tikzpicture}[>={Triangle[bend,width=2pt,length=3pt]}, square/.style={regular polygon,regular polygon sides=4}]
        \coordinate (z);
        \coordinate[right=1 of z] (w);
        \coordinate[right=0.2 of w] (y);  
        \coordinate[above right=0.7 of y] (x1);
        \coordinate[below right=0.7 of y] (x2);
        \node at (x1) [right]{$l^{+}$};
        \node at (x2) [right]{$\nu_{l}$};
        \node at (y)[square,fill,inner sep=1.5pt]{};
        \node at (w)[square,fill,inner sep=1.5pt]{};
        \node at (z)[circle,fill,inner sep=1.3pt]{};      
        \draw[decoration={markings, mark=at position 0.2 with {\coordinate (C);}},postaction={decorate}] (z) [out=45,in=135, looseness=1] to (w);   
  \draw (z) [out=-45,in=-135, looseness=1] to (w);    
        \draw[-{>[sep=15]}] (z) [out=45,in=135, looseness=1] to (w);
        \draw[-{<[sep=15]}] (z) [out=-45,in=-135, looseness=1] to (w);
        \draw (y) to (x1);
        \draw (y) to (x2);
        \draw[-{<[sep=7]}] (y) to (x1);
        \draw[-{>[sep=7]}] (y) to (x2);
        \node at (C)[label={[label distance=-0.2cm]90:$u$}]{};
        \node at (C)[label={[label distance=0.07cm]270:$d$}]{};
        \coordinate[above=0.6 of w] (T1);
        \coordinate[below left = 0.0094 and 0.033 of T1] (T2);
        \node at (T1)[circle,fill,inner sep=0.6pt]{};
        \draw[decoration=Snake,segment length=2.9342574999999pt,segment amplitude=0.9pt, decorate] (T2) arc[start angle=-180,end angle=180,radius=0.2cm];        
        \draw (T1) arc[start angle=0,end angle=360,radius=0.2];
        \draw[->] (T1) arc[start angle=0,end angle=200,radius=0.2];
\end{tikzpicture}}}\ \ \vcenter{\hbox{\begin{tikzpicture}[>={Triangle[bend,width=2pt,length=3pt]}, square/.style={regular polygon,regular polygon sides=4}]
        \coordinate (z);
        \coordinate[right=1 of z] (w);
        \coordinate[right=0.2 of w] (y);
        \coordinate[above right=0.7 of y] (x1);
        \coordinate[below right=0.7 of y] (x2);
        \node at (x1) [right]{$l^{+}$};
        \node at (x2) [right]{$\nu_{l}$};
        \node at (y)[square,fill,inner sep=1.5pt]{};
        \node at (w)[square,fill,inner sep=1.5pt]{};
        \node at (z)[circle,fill,inner sep=1.3pt]{};      
        \draw[decoration={markings, mark=at position 0.2 with {\coordinate (C);}},postaction={decorate}] (z) [out=45,in=135, looseness=1] to (w);   
  \draw (z) [out=-45,in=-135, looseness=1] to (w);    
        \draw[-{>[sep=15]}] (z) [out=45,in=135, looseness=1] to (w);
        \draw[-{<[sep=15]}] (z) [out=-45,in=-135, looseness=1] to (w);
        \draw (y) to (x1);
        \draw (y) to (x2);
        \draw[-{<[sep=7]}] (y) to (x1);
        \draw[-{>[sep=7]}] (y) to (x2);
        \node at (C)[label={[label distance=-0.2cm]90:$u$}]{};
        \node at (C)[label={[label distance=0.07cm]270:$d$}]{};
        \coordinate[above=0.8 of w] (S1);
        \coordinate[below=0.4 of S1] (S2);
        \node at (S2)[circle,fill,inner sep=0.6pt]{};
        \node at (S1)[circle,fill,inner sep=0.6pt]{};      
        \draw (S1) arc[start angle=90,end angle=450,radius=0.2];
        \draw[->] (S1) arc[start angle=90,end angle=200,radius=0.2];
        \draw[->] (S2) arc[start angle=270,end angle=380,radius=0.2];
        \draw[decoration=snake,segment length=2.9pt,segment amplitude=0.9pt,decorate] (S1) -- (S2);  
\end{tikzpicture}}}\ \ \vcenter{\hbox{\begin{tikzpicture}[>={Triangle[bend,width=2pt,length=3pt]}, square/.style={regular polygon,regular polygon sides=4}]
        \coordinate (z);
        \coordinate[right=1 of z] (w);
        \coordinate[right=0.2 of w] (y);  
        \coordinate[above right=0.7 of y] (x1);
        \coordinate[below right=0.7 of y] (x2);
        \node at (x1) [right]{$l^{+}$};
        \node at (x2) [right]{$\nu_{l}$};
        \node at (y)[square,fill,inner sep=1.5pt]{};
        \node at (w)[square,fill,inner sep=1.5pt]{};
        \node at (z)[circle,fill,inner sep=1.3pt]{};      
        \draw[decoration={markings, mark=at position 0.2 with {\coordinate (C);}},postaction={decorate}] (z) [out=45,in=135, looseness=1] to (w);   
  \draw (z) [out=-45,in=-135, looseness=1] to (w);    
        \draw[-{>[sep=15]}] (z) [out=45,in=135, looseness=1] to (w);
        \draw[-{<[sep=15]}] (z) [out=-45,in=-135, looseness=1] to (w);
        \draw (y) to (x1);
        \draw (y) to (x2);
        \draw[-{<[sep=7]}] (y) to (x1);
        \draw[-{>[sep=7]}] (y) to (x2);
        \node at (C)[label={[label distance=-0.2cm]90:$u$}]{};
        \node at (C)[label={[label distance=0.07cm]270:$d$}]{};
        \coordinate[above=0.6 of w] (c);
  \coordinate[left =0.2 of c] (B1);
        \coordinate[right=0.2 of c] (B2);
        \node at (B2)[circle,fill,inner sep=0.6pt]{};
        \node at (B1)[circle,fill,inner sep=0.6pt]{};
        \draw (B1) arc[start angle=0,end angle=360,radius=0.2]
         (B2) arc[start angle=-180,end angle=180,radius=0.2];
        \draw[->] (B1) arc[start angle=0,end angle=200,radius=0.2];
        \draw[->] (B2) arc[start angle=180,end angle=380,radius=0.2];
        \draw[decoration=snake,segment length=2.9pt,segment amplitude=0.9pt,decorate] (B1) -- (B2);        
\end{tikzpicture}}}.
\end{equation}
The renormalized sea quark electromagnetic derivative is computed by cancelling the UV-divergence with the one coming from the scheme observables in the parametrization \eqref{eq:Fpi_par} following the convention from~\cite{Borsanyi:2020mff}.
  
The diagrams are estimated using QED$_{\text{L}}$ in Coulomb gauge and the ensembles listed in Table \ref{tab:sea-ens}.
\begin{table}[!ht]
	\small
    \centering
    \begin{tabular}{c|c|c|c|c|c|c|c}
 action&$\beta$&$a$ [fm]&$L/a \times T/a $&tag& $am_s$&$m_s/m_l$&\#meas\\
 \hline\hline
 \texttt{4stout}&3.7000&0.1315&$24\times 48$&volume/24&0.057291&27.899&716\\
 &&&$48\times 64$&volume/48&0.057291&27.899&300\\
 \hline
 &3.7753&0.1116&$28\times 56$&dir00& 0.047615&27.843&887\\
 \hline
 &3.8400&0.0952&$32\times 64$&dir00&0.043194&28.500&1110\\
 &&&&dir02&0.043194&30.205&1072\\
 &&&&dir04&0.040750&28.007&1036\\
 &&&&dir05&0.039130&26.893&1035\\
 \hline
 \hline
 \texttt{4hex}&0.7300&0.1120&$56\times 84$&phys2/56&0.06061&33.728&1305\\
 \end{tabular}
    \caption{\label{tab:sea-ens} Set of ensembles used for the computation of the sea quark isospin-breaking effects.}
\end{table}

\begin{figure}[!ht]
\centering
\hspace{-0.2cm}\includegraphics[width=11cm]{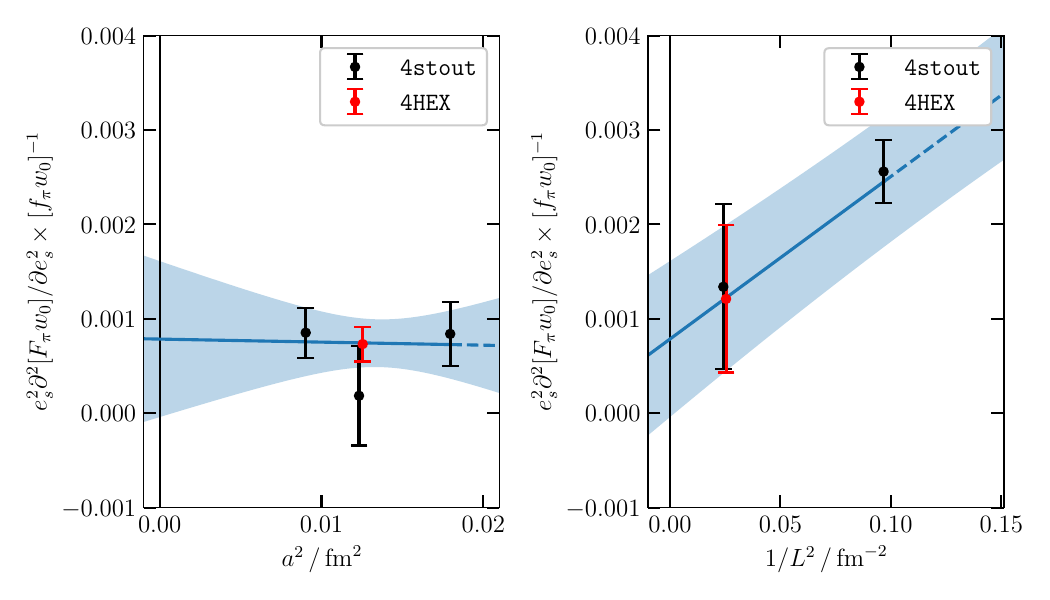}
 \caption{\label{fig:sea_sea}Renormalized sea quark electromagnetic derivative of $w_{0}F_{\pi}$ divided by its QCD value. The left panel shows the continuum, the right panel the infinite volume extrapolation.} 
\end{figure} 

Figure~\ref{fig:sea_sea} shows the effect of the sea derivative relative to the isosymmetric QCD value of $w_{0}F_{\pi}$. As can be seen by the plot, the sea quark IBE in the continuum is about $0.1\%$ of the isosymmetric value, and it shows a flat behaviour in the continuum limit and power-law finite volume effects of order $O\left(1/L^2\right)$ as the leading order structure-dependent finite volume effects~\cite{Lubicz:2016xro}.

The sea-valence IBE, including the following diagrams:
\begin{equation}
\vcenter{\hbox{\begin{tikzpicture}[>={Triangle[bend,width=2pt,length=3pt]}, square/.style={regular polygon,regular polygon sides=4}]
        \coordinate (z);
        \coordinate[right=1 of z] (w);
        \coordinate[above right=0.7 of y] (x1);
        \coordinate[below right=0.7 of y] (x2);
        \node at (x1) [right]{$l^{+}$};
        \node at (x2) [right]{$\nu_{l}$};
        \node at (y)[square,fill,inner sep=1.5pt]{};
        \node at (w)[square,fill,inner sep=1.5pt]{};
        \node at (z)[circle,fill,inner sep=1.3pt]{};      
        \draw[decoration={markings,  mark=at position 0.8 with {\coordinate (B1);}, mark=at position 0.2 with {\coordinate (C);}},postaction={decorate}] (z) [out=45,in=135, looseness=1] to (w);   
  \draw (z) [out=-45,in=-135, looseness=1] to (w);    
        \draw[-{>[sep=15]}] (z) [out=45,in=135, looseness=1] to (w);
        \draw[-{<[sep=15]}] (z) [out=-45,in=-135, looseness=1] to (w);
        \draw (y) to (x1);
        \draw (y) to (x2);
        \draw[-{<[sep=7]}] (y) to (x1);
        \draw[-{>[sep=7]}] (y) to (x2);
        \node at (C)[label={[label distance=-0.2cm]90:$u$}]{};
        \node at (C)[label={[label distance=0.07cm]270:$d$}]{};
        \coordinate[above=0.6 of w] (B2);
        \node at (B2)[circle,fill,inner sep=0.6pt]{};
        \node at (B1)[circle,fill,inner sep=0.6pt]{};
        \draw (B2) arc[start angle=-180,end angle=180,radius=0.2];
        \draw[->] (B2) arc[start angle=180,end angle=380,radius=0.2];
        \draw[decoration=snake,segment length=2.9pt,segment amplitude=0.9pt,decorate] (B1) -- (B2);        
\end{tikzpicture}}}\ \ \vcenter{\hbox{\begin{tikzpicture}[>={Triangle[bend,width=2pt,length=3pt]}, square/.style={regular polygon,regular polygon sides=4}]
        \coordinate (z);
        \coordinate[right=1 of z] (w);
        \coordinate[above right=0.7 of y] (x1);
        \coordinate[below right=0.7 of y] (x2);
        \node at (x1) [right]{$l^{+}$};
        \node at (x2) [right]{$\nu_{l}$};
        \node at (y)[square,fill,inner sep=1.5pt]{};
        \node at (w)[square,fill,inner sep=1.5pt]{};
        \node at (z)[circle,fill,inner sep=1.3pt]{};      
        \draw[decoration={markings, mark=at position 0.2 with {\coordinate (C);}},postaction={decorate}] (z) [out=45,in=135, looseness=1] to (w);   
  \draw[decoration={markings,  mark=at position 0.8 with {\coordinate (B1);}, mark=at position 0.2 with {\coordinate (V);}},postaction={decorate}] (z) [out=-45,in=-135, looseness=1] to (w);    
        \draw[-{>[sep=15]}] (z) [out=45,in=135, looseness=1] to (w);
        \draw[-{<[sep=15]}] (z) [out=-45,in=-135, looseness=1] to (w);
        \draw (y) to (x1);
        \draw (y) to (x2);
        \draw[-{<[sep=7]}] (y) to (x1);
        \draw[-{>[sep=7]}] (y) to (x2);
        \node at (C)[label={[label distance=-0.2cm]90:$u$}]{};
        \node at (C)[label={[label distance=0.07cm]270:$d$}]{};
        \coordinate[below=0.6 of w] (B2);
        \node at (B2)[circle,fill,inner sep=0.6pt]{};
        \node at (B1)[circle,fill,inner sep=0.6pt]{};
        \draw (B2) arc[start angle=-180,end angle=180,radius=0.2];
        \draw[->] (B2) arc[start angle=180,end angle=380,radius=0.2];
        \draw[decoration=snake,segment length=2.9pt,segment amplitude=0.9pt,decorate] (B1) -- (B2);        
\end{tikzpicture}}}\ \ \vcenter{\hbox{\begin{tikzpicture}[>={Triangle[bend,width=2pt,length=3pt]}, square/.style={regular polygon,regular polygon sides=4}]
        \coordinate (z);
        \coordinate[right=1 of z] (w);
        \coordinate[above right=0.7 of y] (x1);
        \coordinate[below right=0.7 of y] (x2);
        \node at (x1) [right]{$l^{+}$};
        \node at (x2) [right]{$\nu_{l}$};
        \node at (y)[square,fill,inner sep=1.5pt]{};
        \node at (w)[square,fill,inner sep=1.5pt]{};
        \node at (z)[circle,fill,inner sep=1.3pt]{};      
        \draw[decoration={markings, mark=at position 0.2 with {\coordinate (C);}},postaction={decorate}] (z) [out=45,in=135, looseness=1] to (w);   
  \draw (z) [out=-45,in=-135, looseness=1] to (w);    
        \draw[-{>[sep=15]}] (z) [out=45,in=135, looseness=1] to (w);
        \draw[-{<[sep=15]}] (z) [out=-45,in=-135, looseness=1] to (w);
        \draw[decoration={markings,mark=at position 0.4 with {\coordinate (B2);}, mark=at position 0.2 with {\coordinate (V);}},postaction={decorate}] (y) to (x1);
        \draw (y) to (x2);
        \draw[-{<[sep=7]}] (y) to (x1);
        \draw[-{>[sep=7]}] (y) to (x2);
        \node at (C)[label={[label distance=-0.2cm]90:$u$}]{};
        \node at (C)[label={[label distance=0.07cm]270:$d$}]{};
        \coordinate[above=0.6 of w] (B1);
        \node at (B1)[circle,fill,inner sep=0.6pt]{};
        \node at (B2)[circle,fill,inner sep=0.6pt]{};
        \draw (B1) arc[start angle=0,end angle=360,radius=0.2];
        \draw[->] (B1) arc[start angle=0,end angle=200,radius=0.2];
        \draw[decoration=snake,segment length=2.9pt,segment amplitude=0.9pt,decorate] (B1) -- (B2);        
\end{tikzpicture}}},
\end{equation}
are estimated to be of the order of $20\%$ compared to the sea-sea derivative and thus neglected in our computation.

\section{Total Error Budget}
Combining the results for the isoQCD and the IBE as mentioned in Section~\ref{sec:2_mixed_det} we obtain the following preliminary QCD+QED value of $w_{0}F_{\pi}$:
\begin{equation}
\left[w_{0}F_{\pi}\right]_{\text{QCD+QED}} = 0.11527(15)(26)[31]\ \text{(Preliminary)},
\end{equation}
where the first error is the statistical, the second is systematic, and the latter is the combined total error. The systematic error consists of the following elements: 
\begin{equation}
(26)=(23)_{\text{QCD}}(11)_{\text{QED-Valence}}(5)_{\text{QED-Sea}}\ \text{(Preliminary)}.
\end{equation}
The QCD error dominates the final uncertainty, while the QED error is dominated by the valence QED value.

Combining $w_{0}F_{\pi}$ with the QCD+QED value of $F_{\pi}=131.711(45)$ MeV (obtained using $V_{ud} =0.97367(32)$ and $\Gamma = 3.8408(7)\times 10^{7}$ s$^{-1}$ from~\cite{ParticleDataGroup:2022pth}) we get the following value of $w_{0}$:
\begin{equation}
\left[w_{0}\right]_{\text{QCD+QED}} = 0.17270(22)(40)[46]\ \text{fm}\ \text{(Preliminary)}.
\end{equation}

The systematic error is made by the following contributions:
\begin{equation}
(40)=(34)_{\text{QCD}}(17)_{\text{QED-Valence}}(8)_{\text{QED-Sea}}(7)_{\text{exp}}\ \text{(Preliminary)},
\end{equation} 
where the additional systematic error \textit{exp} accounts for the experimental value of $F_{\pi}$. We are working on improving the QCD error and improving the valence QED determination by providing an independent determination of $\delta R_{\pi}$ and by checking the power-law finite volume effects in light of the work from~\cite{DiCarlo:2021apt}.

\section{Ongoing Work on the Valence Quark Isospin-Breaking Effects}
The valence-quark IBE in the pion decay constant requires the computation of the non-factorizable diagrams:
\begin{equation}
\vcenter{\hbox{\begin{tikzpicture}[>={Triangle[bend,width=2pt,length=3pt]}, square/.style={regular polygon,regular polygon sides=4}]
        \coordinate (z);
        \coordinate[right=1 of z] (w);
        \coordinate[above right=0.7 of y] (x1);
        \coordinate[below right=0.7 of y] (x2);
        \node at (x1) [right]{$l^{+}$};
        \node at (x2) [right]{$\nu_{l}$};
        \node at (y)[square,fill,inner sep=1.5pt]{};
        \node at (w)[square,fill,inner sep=1.5pt]{};
        \node at (z)[circle,fill,inner sep=1.3pt]{};      
        \draw[decoration={markings,  mark=at position 0.8 with {\coordinate (B1);}, mark=at position 0.2 with {\coordinate (C);}},postaction={decorate}] (z) [out=45,in=135, looseness=1] to (w);   
  \draw (z) [out=-45,in=-135, looseness=1] to (w);    
        \draw[-{>[sep=15]}] (z) [out=45,in=135, looseness=1] to (w);
        \draw[-{<[sep=15]}] (z) [out=-45,in=-135, looseness=1] to (w);
        \draw[decoration={markings,mark=at position 0.4 with {\coordinate (B2);}, mark=at position 0.2 with {\coordinate (V);}},postaction={decorate}] (y) to (x1);
        \draw (y) to (x2);
        \draw[-{<[sep=7]}] (y) to (x1);
        \draw[-{>[sep=7]}] (y) to (x2);
        \node at (B1)[circle,fill,inner sep=0.6pt]{};      
        \node at (B2)[circle,fill,inner sep=0.6pt]{};              
        \node at (C)[label={[label distance=-0.2cm]90:$u$}]{};
        \node at (C)[label={[label distance=0.07cm]270:$d$}]{};
        \draw[decoration=snake,segment length=3.3pt,segment amplitude=0.8pt,decorate] (B1) to [bend left=50] (B2);        
\end{tikzpicture}}}\ \ \vcenter{\hbox{\begin{tikzpicture}[>={Triangle[bend,width=2pt,length=3pt]}, square/.style={regular polygon,regular polygon sides=4}]
        \coordinate (z);
        \coordinate[right=1 of z] (w);
        \coordinate[above right=0.7 of y] (x1);
        \coordinate[below right=0.7 of y] (x2);
        \node at (x1) [right]{$l^{+}$};
        \node at (x2) [right]{$\nu_{l}$};
        \node at (y)[square,fill,inner sep=1.5pt]{};
        \node at (w)[square,fill,inner sep=1.5pt]{};
        \node at (z)[circle,fill,inner sep=1.3pt]{};      
        \draw[decoration={markings, mark=at position 0.2 with {\coordinate (C);}},postaction={decorate}] (z) [out=45,in=135, looseness=1] to (w);   
  \draw[decoration={markings,  mark=at position 0.7 with {\coordinate (B1);}, mark=at position 0.2 with {\coordinate (V);}},postaction={decorate}] (z) [out=-45,in=-135, looseness=1] to (w);    
        \draw[-{>[sep=15]}] (z) [out=45,in=135, looseness=1] to (w);
        \draw[-{<[sep=15]}] (z) [out=-45,in=-135, looseness=1] to (w);
\draw[decoration={markings,mark=at position 0.4 with {\coordinate (B2);}, mark=at position 0.2 with {\coordinate (V);}},postaction={decorate}] (y) to (x1);
        \draw (y) to (x2);
        \draw (y) to (x2);
        \draw[-{<[sep=7]}] (y) to (x1);
        \draw[-{>[sep=7]}] (y) to (x2);
        \node at (C)[label={[label distance=-0.2cm]90:$u$}]{};
        \node at (C)[label={[label distance=0.07cm]270:$d$}]{};
        \node at (B2)[circle,fill,inner sep=0.6pt]{};
        \node at (B1)[circle,fill,inner sep=0.6pt]{};
        \draw[decoration=snake,segment length=3.5pt,segment amplitude=0.9pt,decorate] (B1) to [bend left=70] (B2);        
\end{tikzpicture}}}
\end{equation}
 (actively involving the lepton) and the factorizable ones; we present some preliminary results of the factorizable diagrams.
The factorizable valence contributions to the electromagnetic derivative of the pion decay constant are given by the following diagrams:
\begin{equation}
\vcenter{\hbox{\begin{tikzpicture}[>={Triangle[bend,width=2pt,length=3pt]}, square/.style={regular polygon,regular polygon sides=4}]
        \coordinate (z);
        \coordinate[right=1 of z] (w);
        \coordinate[right=0.2 of w] (y);  
        \coordinate[above right=0.7 of y] (x1);
        \coordinate[below right=0.7 of y] (x2);
        \node at (x1) [right]{$l^{+}$};
        \node at (x2) [right]{$\nu_{l}$};
        \node at (y)[square,fill,inner sep=1.5pt]{};
        \node at (w)[square,fill,inner sep=1.5pt]{};
        \node at (z)[circle,fill,inner sep=1.3pt]{}; 
  \draw[decoration={markings, mark=at position 0.2 with {\coordinate (F);}},postaction={decorate}] (z) [out=45,in=135, looseness=1] to (w);         
        \draw[decoration={markings, mark=at position 0.70 with {\coordinate (B);},mark=at position 0.25 with {\coordinate (E);}},postaction={decorate}] (z) [out=45,in=135, looseness=1] to (w);
          \node at (B)[circle,fill,inner sep=0.6pt]{};
  \draw[decoration={markings, mark=at position 0.30 with {\coordinate (C);},mark=at position 0.75 with {\coordinate (D);}
  },postaction={decorate}] (z) [out=-45,in=-135, looseness=1] to (w);    
       \node at (C)[circle,fill,inner sep=0.6pt]{};      
       \draw[-{>[sep=15]}] (z) [out=45,in=135, looseness=1] to (w);
        \draw[-{<[sep=15]}] (z) [out=-45,in=-135, looseness=1] to (w);
                \draw[decoration=snake,segment length=2.9pt,segment amplitude=0.9pt,decorate] (B) -- (C);
        \draw (y) to (x1);
        \draw (y) to (x2);
        \draw[-{<[sep=7]}] (y) to (x1);
        \draw[-{>[sep=7]}] (y) to (x2);
        \node at (F)[label={[label distance=-0.2cm]90:$u$}]{};
        \node at (F)[label={[label distance=0.07cm]270:$d$}]{};
\end{tikzpicture}}}\vcenter{\hbox{\begin{tikzpicture}[>={Triangle[bend,width=2pt,length=3pt]}, square/.style={regular polygon,regular polygon sides=4}]
        \coordinate (z);
        \coordinate[right=1 of z] (w);
        \coordinate[right=0.2 of w] (y);  
  \coordinate[above right=0.7 of y] (x1);
        \coordinate[below right=0.7 of y] (x2);
        \node at (x1) [right]{$l^{+}$};
        \node at (x2) [right]{$\nu_{l}$};
        \node at (y)[square,fill,inner sep=1.5pt]{};
        \node at (w)[square,fill,inner sep=1.5pt]{};
        \node at (z)[circle,fill,inner sep=1.3pt]{};  
        \draw[decoration={markings, mark=at position 0.8 with {\coordinate (B);}
  },postaction={decorate}] (z) [out=45,in=135, looseness=1] to (w);
          \draw[decoration={markings, mark=at position 0.2 with {\coordinate (C);}
  },postaction={decorate}] (z) [out=45,in=135, looseness=1] to (w);   
     \node at (B)[circle,fill,inner sep=0.6pt]{};
 \draw[decoration={markings, mark=at position 0.75 with {\coordinate (D);}
  },postaction={decorate}] (z) [out=-45,in=-135, looseness=1] to (w);    
          \node at (C)[circle,fill,inner sep=0.6pt]{};      
        \draw[-{>[sep=15]}] (z) [out=45,in=135, looseness=1] to (w);
        \draw[-{<[sep=15]}] (z) [out=-45,in=-135, looseness=1] to (w);
        \draw[decoration=snake,segment length=2.9pt,segment amplitude=0.9pt,decorate] (C)  to [bend right=50] (B); 
         \draw (y) to (x1);
        \draw (y) to (x2);
        \draw[-{<[sep=7]}] (y) to (x1);
        \draw[-{>[sep=7]}] (y) to (x2);
        \node at (C)[label={[label distance=-0.2cm]90:$u$}]{};
        \node at (C)[label={[label distance=0.07cm]270:$d$}]{};
\end{tikzpicture}}}\vcenter{\hbox{\begin{tikzpicture}[>={Triangle[bend,width=2pt,length=3pt]}, square/.style={regular polygon,regular polygon sides=4}]
        \coordinate (z);
        \coordinate[right=1 of z] (w);
        \coordinate[right=0.2 of w] (y);  
  \coordinate[above right=0.7 of y] (x1);
        \coordinate[below right=0.7 of y] (x2);
        \node at (x1) [right]{$l^{+}$};
        \node at (x2) [right]{$\nu_{l}$};
        \node at (y)[square,fill,inner sep=1.5pt]{};
        \node at (w)[square,fill,inner sep=1.5pt]{};
        \node at (z)[circle,fill,inner sep=1.3pt]{}; 
  \draw[decoration={markings, mark=at position 0.2 with {\coordinate (F);}},postaction={decorate}] (z) [out=45,in=135, looseness=1] to (w);
        \draw[decoration={markings, mark=at position 0.8 with {\coordinate (B);}
  },postaction={decorate}] (z) [out=-45,in=-135, looseness=1] to (w);
          \draw[decoration={markings, mark=at position 0.2 with {\coordinate (C);}
  },postaction={decorate}] (z) [out=-45,in=-135, looseness=1] to (w);   
     \node at (B)[circle,fill,inner sep=0.6pt]{};
     \node at (C)[circle,fill,inner sep=0.6pt]{};      
        \draw[-{>[sep=15]}] (z) [out=45,in=135, looseness=1] to (w);
        \draw[-{<[sep=15]}] (z) [out=-45,in=-135, looseness=1] to (w);        
        \draw[decoration=snake,segment length=2.7pt,segment amplitude=0.9pt,decorate] (C)  to [bend left=50] (B);    
        \draw (y) to (x1);
        \draw (y) to (x2);
        \draw[-{<[sep=7]}] (y) to (x1);
        \draw[-{>[sep=7]}] (y) to (x2);
        \node at (F)[label={[label distance=-0.2cm]90:$u$}]{};
        \node at (F)[label={[label distance=0.07cm]270:$d$}]{};
\end{tikzpicture}}}\vcenter{\hbox{\begin{tikzpicture}[>={Triangle[bend,width=2pt,length=3pt]}, square/.style={regular polygon,regular polygon sides=4}]
  \coordinate (z);
        \coordinate[right=1 of z] (w);
        \coordinate[right=0.2 of w] (y);  
        \coordinate[above right=0.7 of y] (x1);
        \coordinate[below right=0.7 of y] (x2);
        \node at (x1) [right]{$l^{+}$};
        \node at (x2) [right]{$\nu_{l}$};
        \node at (y)[square,fill,inner sep=1.5pt]{};
        \node at (w)[square,fill,inner sep=1.5pt]{};
        \node at (z)[circle,fill,inner sep=1.3pt]{}; 
        \draw[decoration={markings, mark=at position 0.8 with {\coordinate (B);},mark=at position 0.2 with {\coordinate (C);}},postaction={decorate}] (z) [out=45,in=135, looseness=1] to (w);    
        \node at (B)[circle,fill,inner sep=0.6pt]{};
  \draw[decoration={markings,mark=at position 0.75 with {\coordinate (D);}},postaction={decorate}]  (z) [out=-45,in=-135, looseness=1] to (w);
  \draw[decoration=Snake,segment length=2.9342574999999pt,segment amplitude=0.9pt, decorate] (B) arc[start angle=-90,end angle=270,radius=0.2cm];          
        \draw[-{>[sep=15]}] (z) [out=45,in=135, looseness=1] to (w);
        \draw[-{<[sep=15]}] (z) [out=-45,in=-135, looseness=1] to (w);
        \draw (y) to (x1);
        \draw (y) to (x2);
        \draw[-{<[sep=7]}] (y) to (x1);
        \draw[-{>[sep=7]}] (y) to (x2);
        \node at (C)[label={[label distance=-0.2cm]90:$u$}]{};
        \node at (C)[label={[label distance=0.07cm]270:$d$}]{};
        \end{tikzpicture}}}\ 
\vcenter{\hbox{\begin{tikzpicture}[>={Triangle[bend,width=2pt,length=3pt]}, square/.style={regular polygon,regular polygon sides=4}]
  \coordinate (z);
        \coordinate[right=1 of z] (w);
        \coordinate[right=0.2 of w] (y);  
        \coordinate[above right=0.7 of y] (x1);
        \coordinate[below right=0.7 of y] (x2);
        \node at (x1) [right]{$l^{+}$};
        \node at (x2) [right]{$\nu_{l}$};
        \node at (y)[square,fill,inner sep=1.5pt]{};
        \node at (w)[square,fill,inner sep=1.5pt]{};
        \node at (z)[circle,fill,inner sep=1.3pt]{}; 
        \draw[decoration={markings, mark=at position 0.2 with {\coordinate (C);}},postaction={decorate}] (z) [out=45,in=135, looseness=1] to (w);
  \draw[decoration={markings,mark=at position 0.8 with {\coordinate (B);}},postaction={decorate}]  (z) [out=-45,in=-135, looseness=1] to (w);
        \node at (B)[circle,fill,inner sep=0.6pt]{};
  \draw[decoration=Snake,segment length=2.9342574999999pt,segment amplitude=0.9pt, decorate] (B) arc[start angle=-270,end angle=90,radius=0.2cm];          
         \draw[-{>[sep=15]}] (z) [out=45,in=135, looseness=1] to (w);
        \draw[-{<[sep=15]}] (z) [out=-45,in=-135, looseness=1] to (w);
        \draw (y) to (x1);
        \draw (y) to (x2);
        \draw[-{<[sep=7]}] (y) to (x1);
        \draw[-{>[sep=7]}] (y) to (x2);
        \node at (C)[label={[label distance=-0.2cm]90:$u$}]{};
        \node at (C)[label={[label distance=0.07cm]270:$d$}]{};
\end{tikzpicture}}}.
\end{equation}
The diagrams are computed using sequential propagators and $16$ photon sources in Coulomb gauge and $3$ U(1) random sources. The zeroth mode of the photon is removed following the QED$_{\text{L}}$ prescription. We use the ensembles listed in Table~\ref{tab:valence-ens}. We tested the program by comparing it with numerical derivatives at a fixed random source and photon source. 

The correction is obtained by expanding the axial-pseudoscalar (using the conserved axial current) and pseudoscalar-pseudoscalar two-point function:
\begin{equation}
\frac{G^{PA_{4}}_{ud}(x_{4})^2}{G^{PP}_{ud}(x_{4})} = M_{\pi}Z_{A}^{-2}F_{\pi}^2e^{-M_{\pi}T/2}\frac{\sinh^2\left[M_{\pi}\left(T/2-x_{4}-1/2\right)\right]}{\cosh\left[M_{\pi}\left(T/2-x_{4}\right)\right]}
\end{equation}
where $T$ is the temporal exension and $Z_{A}$ is the renormalization constant which is not considered in these results and, thus, it is left for future work.
\begin{table}[!ht]
\small
    \centering
\begin{tabular}{c|c|c|c|c|c|c|c}
 action&$\beta$&$a$ [fm]&$L/a \times T/a $&tag& $am_s$&$m_s/m_l$&\#confs\\
 \hline\hline
 \texttt{4stout}&3.7000&0.1315&$24\times 48$&volume/24&0.057291&27.899&48\\
 &&&$32\times 64$&volume/32&0.057291&27.899&48\\ 
 &&&$48\times 64$&volume/48&0.057291&27.899&48\\
 &&&$64\times 64$&volume/64&0.057291&27.899&48\\ 
 \hline
 &3.8400&0.0952&$64\times 96$&dir00&0.043194&28.500&48\\
    \end{tabular}
\caption{\label{tab:valence-ens} Set of ensembles used for the computation of the valence quarks isospin-breaking effects.}
\end{table}

\begin{figure}[!ht]
  \centering
  \includegraphics[width=9cm]{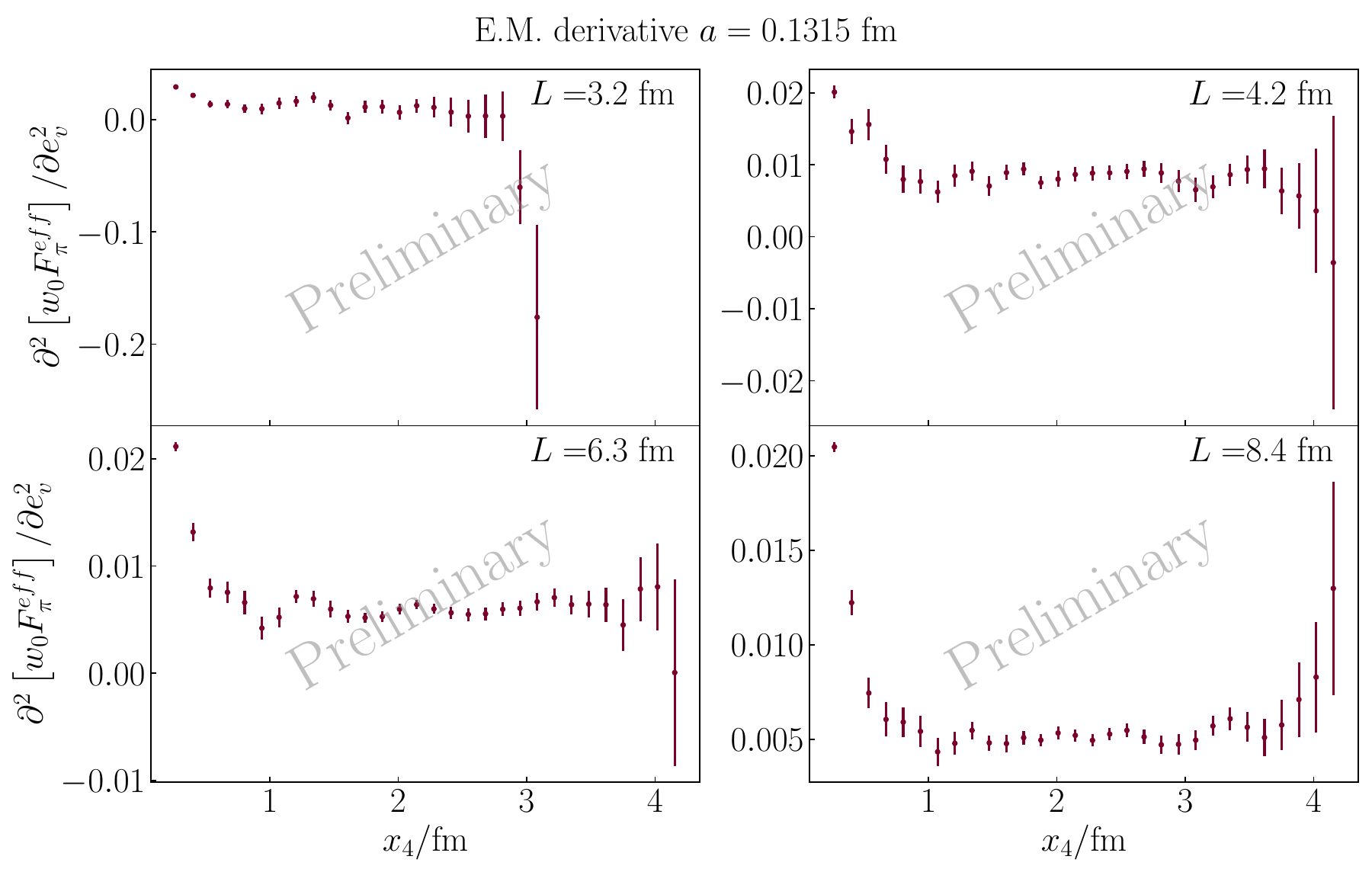} 
  \caption{\label{fig:fact_val_beta3.7}Factorizable valence electromagnetic derivative of $w_{0}F_{\pi}$ for $a=0.1315$ fm. The preliminary results for the four volumes are shown.}
\end{figure}

Figure~\ref{fig:fact_val_beta3.7} shows the preliminary result for the effective factorizable valence derivative for the four different volumes considered.

\begin{figure}[!ht]
  \centering
  \includegraphics[width=7.9cm]{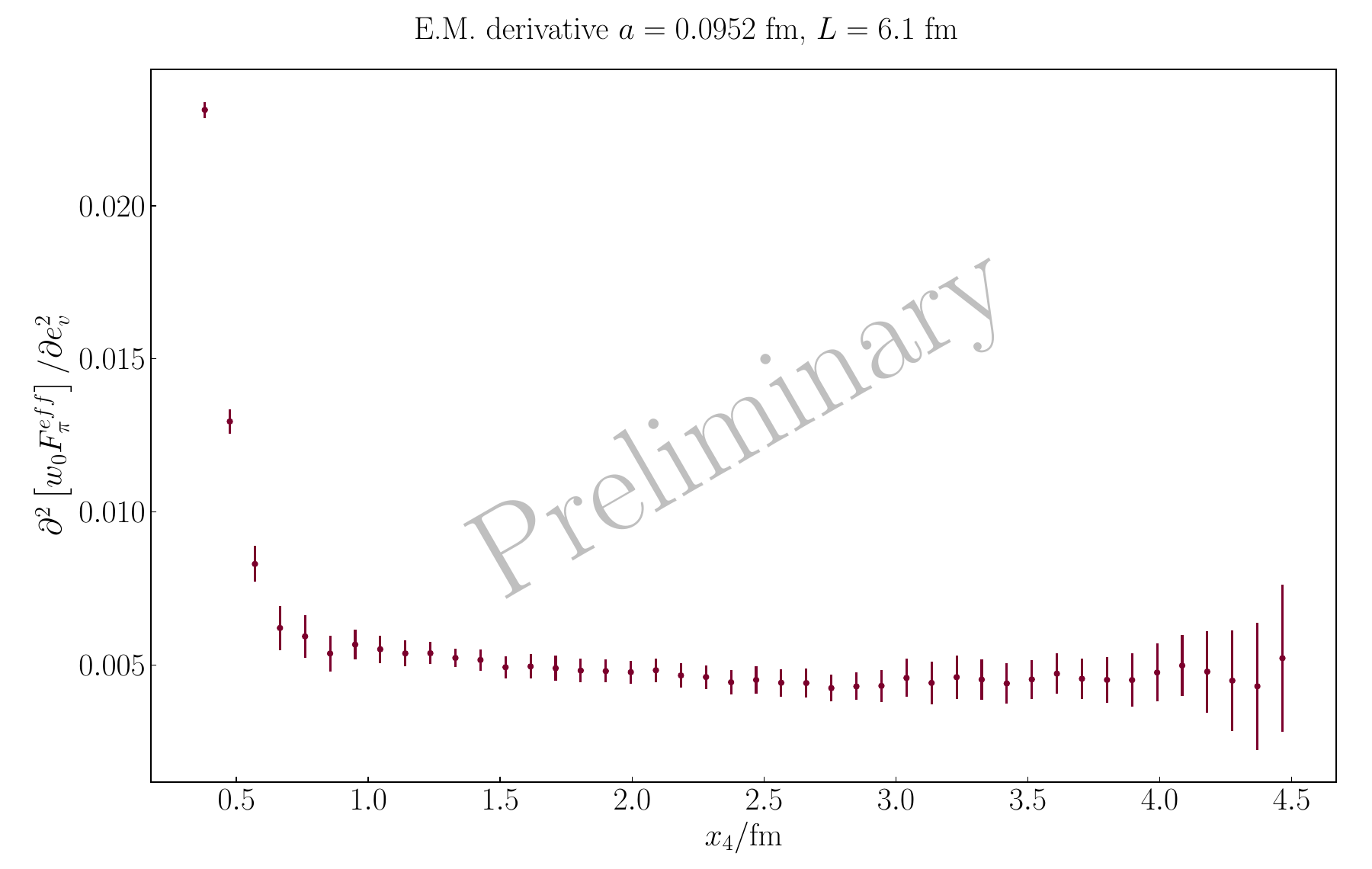} 
    \caption{\label{fig:fact_val_beta3.84}Factorizable valence electromagnetic derivative of $w_{0}F_{\pi}$ for $a=0.0952$ fm.}
\end{figure}

Figure~\ref{fig:fact_val_beta3.84} shows the preliminary result for the effective factorizable valence derivative for the finer value of the lattice spacing. The effective derivative presents a clear plateau at large time separation, showing little excited states contamination and little oscillating contribution.

\section{Conclusions \& Outlook}
In this Proceedings, we have presented the results of the current determination of the pion decay constant in QCD+QED by the BMW collaboration for the determination of the scale setting quantity $w_{0}$, the preliminary value is:
\begin{equation}
\left[w_{0}\right]_{\text{QCD+QED}} = 0.17270(22)(40)[46]\ \text{fm}\ \text{(Preliminary)},
\end{equation}
where the precision is dominated by the continuum extrapolation of the isosymmetric part. The determination of isospin-breaking effects was done by combining sea quark effects from BMW ensembles and the valence-quark effects from the RM123 work. 

We are working on improving the continuum extrapolation (by adding a finer lattice spacing) and on a completely independent determination of the valence effects, which will include the sea-valence (that have been estimated to be suppressed). This requires the determination of the non-factorizable diagrams involving the leptons and the renormalization of the electroweak Hamiltonian. The authors of~\cite{DiCarlo:2021apt,Boyle:2022lsi} showed the poor convergence of the series of finite volume effects to the pion decay constant in QED$_{\text{L}}$, for this reason, we plan a detailed study of the finite volume effects using different volumes (up to $10.8$ fm).

\acknowledgments
DG thanks Nazario Tantalo for useful discussions on the GRS scheme. The authors gratefully acknowledge the Gauss Centre for Supercomputing (GCS) e.V. ({\tt www.gauss-centre.eu}),
GENCI ({\tt www.genci.fr}, grant 502275), EuroHPC Joint Undertaking
(grants EXT-2023E02-063, EXT-2024E02-109) and Australian National Computational Merit Allocation
Scheme for providing computer time on the GCS supercomputers SuperMUC-NG at
Leibniz Supercomputing Centre in München, HAWK and HUNTER at the High Performance
Computing Center in Stuttgart and JUWELS, JURECA and JUPITER at Forschungszentrum
Jülich, as well as on the GENCI supercomputers Joliot-Curie/Irène Rome at TGCC,
Jean-Zay V100 at IDRIS, Adastra at CINES, on the EuroHPC JU flagship supercomputers Leonardo
at CINECA and LUMI at CSC, Gadi at NCI and Setonix at PSC. DG is supported by ERC-MUON-101054515.
\bibliographystyle{JHEP}
\bibliography{inspire}

\end{document}